\documentclass[twocolumn]{aastex631}
\usepackage{graphicx}
\usepackage{soul}
\usepackage{xurl}
\usepackage[utf8]{inputenc}
\usepackage{url}
\usepackage{booktabs}

\usepackage{xcolor}

\usepackage[normalem]{ulem}

\begin{document}

\title{A Shallow Phase-resolved Gamma-Ray Luminosity-Cutoff Energy Relation in Globular Cluster Millisecond Pulsars}
%\title{A Phase-resolved \textit{Fermi}-LAT Study of High-energy Emission from Four Globular Cluster Millisecond Pulsars}

\author{Mengqing Zhang}
\affiliation{Department of Astronomy, School of Physics and Astronomy, Key Laboratory of Astroparticle Physics of Yunnan Province, Yunnan University, Kunming 650091, People's Republic of China; changshan@ynu.edu.cn, zhangpengfei@ynu.edu.cn}

\author{Jiandong Zhang}
\affiliation{Department of Astronomy, School of Physics and Astronomy, Key Laboratory of Astroparticle Physics of Yunnan Province, Yunnan University, Kunming 650091, People's Republic of China; changshan@ynu.edu.cn, zhangpengfei@ynu.edu.cn}

\author{Shan Chang}
\affiliation{Department of Astronomy, School of Physics and Astronomy, Key Laboratory of Astroparticle Physics of Yunnan Province, Yunnan University, Kunming 650091, People's Republic of China; changshan@ynu.edu.cn, zhangpengfei@ynu.edu.cn}

\author{Pengfei Zhang}
%\correspondingauthor{Pengfei Zhang}
\affiliation{Department of Astronomy, School of Physics and Astronomy, Key Laboratory of Astroparticle Physics of Yunnan Province, Yunnan University, Kunming 650091, People's Republic of China; changshan@ynu.edu.cn, zhangpengfei@ynu.edu.cn}

\begin{abstract}

Globular clusters (GCs) provide dense stellar environments for investigating
the high-energy emission of millisecond pulsars (MSPs). Using $\sim$17.3~yr
of \textit{Fermi}-LAT Pass~8 data in the 0.1--500.0~GeV range, we study
four MSPs that dominate, or contribute substantially to, the GeV $\gamma$-ray
emission of their host GCs. Significant off-pulse $\gamma$-ray emission is
detected from PSR~J1823$-$3021A, J1824$-$2452A, and J1835$-$3259B.
For these sources, the cutoff energy, $E_{\rm cut}$, varies systematically
with rotational phase and is positively correlated with the phase-resolved
energy flux. Using the phase-resolved isotropic $\gamma$-ray luminosity
($L_{\gamma}$), we find $L_{\gamma}\propto E_{\rm cut}^{\alpha}$ with
$\alpha=0.74^{+0.16}_{-0.14}$. This slope is substantially shallower than
that reported for field MSPs, $\alpha=2.31^{+0.22}_{-0.25}$,
and the $\alpha=16/7\simeq2.29$ scaling expected for curvature radiation
in the equatorial current sheet framework. The origin of the shallow
relation remains uncertain, potentially reflecting cluster-associated
$\gamma$-ray emission, differences in emission geometry and beaming,
or limited photon statistics. Our results demonstrate that GC MSPs
provide valuable targets for testing phase-resolved high-energy emission
physics in dense stellar environments.

\end{abstract}


\section{Introduction}

Millisecond pulsars (MSPs) are established GeV $\gamma$-ray emitters whose high-energy
radiation provides a sensitive probe of particle acceleration and emission geometry in
pulsar magnetospheres. Globular clusters (GCs), with their high stellar densities and
frequent dynamical interactions, are particularly efficient environments for forming
and retaining MSPs and therefore provide valuable laboratories for investigating MSP
high-energy emission in dense stellar systems. Since the launch of the \textit{Fermi}
Gamma-ray Space Telescope in 2008 and the operation of its Large Area Telescope (LAT;
\citealt{atw+09}), several dozen GCs have been detected as GeV $\gamma$-ray sources
\citep{abd+10,aaa+10,tam+11,zzh+15,zxf+16,lcb18,dcn19,abd+20,yrz+22,yzz+22,aab+22,wwx+22}.
Their $\gamma$-ray emission is commonly interpreted as the integrated contribution of
an unresolved MSP population within the cluster. In a small number of systems, however,
timing and spectral studies suggest that a single identified MSP dominates, or makes
a substantial contribution to, the cluster $\gamma$-ray
emission \citep{fre+11,whw+13,joh+13,zxw22,zxw+23,zzz24}. Such systems provide rare
opportunities to isolate the high-energy emission of individual MSPs in GC environments
and to assess the potential influence of the cluster environment on their observed
$\gamma$-ray properties.

Despite the wealth of high-energy observations now available for MSPs, the location
and physical mechanism of their $\gamma$-ray emission remain under active investigation.
In conventional magnetospheric models, particles are accelerated by rotation-induced
electric fields in regions of incomplete charge screening within the light cylinder (LC),
and the resulting high-energy emission is commonly attributed primarily to curvature
radiation. Such models include the outer-gap (OG) model \citep{chr86,rom96,zc97},
the slot-gap (SG) and two-pole-caustic (TPC) models \citep{aro83,mh04,dr03}, and
the pair-starved polar-cap (PSPC) model \citep{mh04}. More recently, global magnetospheric
simulations have shown that the equatorial current sheet (ECS), which forms near and
beyond the LC, may provide an additional particle-acceleration site, where particles
emit radiation via curvature radiation, synchrotron radiation, or a combination of
both \citep{bs10, cps16, ps18}. The interpretation is further
complicated for MSPs hosted by GCs, where the observed $\gamma$-ray emission may
include contributions from processes outside the pulsar magnetosphere. Relativistic
electrons and positrons escaping from pulsar winds can inverse-Compton (IC) scatter
the enhanced optical and infrared radiation fields produced by cluster stars \citep{hl18, slb21},
while MSPs in binary systems may additionally produce high-energy emission in
intrabinary shock (IBS) regions \citep{acr+82, ack+10}. Distinguishing magnetospheric
emission from such cluster-associated or binary-related components is therefore
essential for establishing the origin of the $\gamma$-ray radiation from individual
GC MSPs.

Phase-resolved observations provide important constraints on MSP $\gamma$-ray emission
models because they probe how the emission intensity and spectral shape vary across
the rotational cycle. In particular, emission detected away from the main pulse peaks
(hereafter off-pulse emission) provides a useful diagnostic of both the extent of
magnetospheric radiation and possible additional unpulsed components.
Very weak off-pulse emission may be consistent with radiation concentrated in narrow
caustic beams, whereas significant off-pulse emission can arise from
broad magnetospheric beams or extended acceleration regions, but may also include
a steady contribution from processes outside the magnetosphere. The off-pulse spectral
shape offers a complementary, though not definitive, diagnostic: a GeV cutoff spectrum
is generally consistent with magnetospheric curvature radiation, whereas
an approximately power-law component may suggest contributions from IC scattering,
pulsar-wind nebulae, or IBSs \citep{ack+10, aaa+13}.
Recently, \citet{lsx+26} performed a systematic phase-resolved analysis of 38 MSPs
in the Galactic field (hereafter field MSPs), primarily drawn from pulsar timing array
samples, and detected significant off-pulse emission from 15 sources.
Ten of these sources showed significant spectral cutoffs, 
and 11 of these 15 sources exhibited a significant positive correlation between the phase-resolved
cutoff energy ($E_{\rm cut}$) and photon counts.
Using a phase-resolved pseudo-luminosity,
\citet{lsx+26} further found a relation of $L_{\gamma}\propto E_{\rm cut}^{\alpha}$ with
$\alpha=2.31^{+0.22}_{-0.25}$, consistent with the curvature-radiation scaling expected
in the ECS framework. These results provide an observational benchmark for assessing
whether MSPs in GC environments show comparable phase-resolved spectral behavior and
luminosity--$E_{\rm cut}$ relations.

Motivated by the phase-resolved results obtained for field MSPs, we extend this investigation
to four MSPs hosted by GCs: PSR~J1823$-$3021A, PSR~J1824$-$2452A, PSR~J1835$-$3259B, and
PSR~J1717$+$4308A. Previous timing and spectral studies suggest that these pulsars either
dominate, or make substantial contributions to, the GeV $\gamma$-ray emission of their
respective host clusters \citep{fre+11, whw+13, joh+13, zxw22, zxw+23}. Using high-precision
timing solutions, including an updated ephemeris derived for PSR~J1824$-$2452A, we fold
the \textit{Fermi}-LAT photons to construct probability-weighted $\gamma$-ray pulse profiles,
define off-pulse phase intervals, and perform off-pulse and phase-resolved spectral analyses.
Significant off-pulse emission is detected from PSR~J1823$-$3021A, PSR~J1824$-$2452A,
and PSR~J1835$-$3259B. For these three sources, the phase-resolved energy flux and $E_{\rm cut}$
are positively correlated; however, the combined phase-resolved $L_{\gamma}$--$E_{\rm cut}$
relation has a slope of $\alpha=0.74^{+0.16}_{-0.14}$, substantially shallower than the
$\alpha=2.31^{+0.22}_{-0.25}$ relation reported for field MSPs by \citet{lsx+26}. Thus,
although the detection of off-pulse emission and phase-dependent spectral evolution supports
a magnetospheric contribution over a broad range of rotational phases, the different
$L_{\gamma}$--$E_{\rm cut}$ scaling suggests that GC MSPs may not follow the same simple
phase-resolved relation inferred for field MSPs. The data selection and analysis procedures
are described in Section~\ref{sec:methods}, and the results and their implications are presented
in the subsequent Section~\ref{sec:summary_discussion}.

\section{Data analysis and results}
\label{sec:methods}

\subsection{Sample Selection and Data Reduction}
\label{sec:data_selection}

The targets of this study are MSPs hosted by GCs, whose GeV $\gamma$-ray emission
has been detected by \textit{Fermi}-LAT and is thought to be dominated by, or to
include a substantial contribution from, at least one confirmed MSP. We selected
the sample according to the following criteria: (1) the host GC is detected as a
GeV $\gamma$-ray source by \textit{Fermi}-LAT; (2) previous studies indicate that
its $\gamma$-ray emission is likely dominated by a single confirmed MSP or that
this MSP contributes significantly to the total cluster emission, rather than
arising solely from the unresolved collective emission of the MSP population; and
(3) the target MSP has a sufficiently precise and sufficiently long timing solution
to enable reliable phase assignment of \textit{Fermi}-LAT photons and subsequent
analyses of its pulsed and off-pulse emission, as well as its phase-resolved spectra.

\begin{table*}
\centering
\caption{Information for the four GC MSPs}
\label{tab:msp_parameters}
\renewcommand{\arraystretch}{1.2}
\begin{tabular}{lcccccc}
\hline\hline
 & & \multicolumn{2}{c}{Coordinate} & & & \\
\cline{3-4}
\multicolumn{1}{c}{Pulsar name} & Host GC & R.A. & Dec. & $D$ (kpc) & $P$ (ms) & Binary \\
\hline
J1823$-$3021A & NGC 6624 & 275.892 & $-$30.350 & $7.60 \pm 0.40$$^a$ & 5.44 & N \\
J1824$-$2452A & NGC 6626 & 276.161 & $-$24.870 & $5.37 \pm 0.10$$^a$ & 3.05 & N \\
J1835$-$3259B & NGC 6652 & 278.940 & $-$32.971 & $8.40^{+0.94}_{-0.77}$$^a$ & 1.83 & Y \\
J1717$+$4308A & NGC 6341 & 259.203 & $+$43.173 & $8.50 \pm 0.07$$^b$ & 3.16 & Y \\
\hline
\end{tabular}
\begin{flushleft} 
\small
\tablecomments{Distance $D$ to each host GC is given in kpc. Distances marked with $^{a}$
                are adopted primarily from \citet{smi+23}, whereas the distance marked with
                $^{b}$ is taken from \citet{bv21}. The rotation period $P$ of each MSP is
                given in milliseconds. The column ``Binary'' indicates whether the MSP is
                in a binary system (Y: yes; N: no).}
\end{flushleft}
\end{table*}

Based on these criteria, we selected four MSPs hosted by GCs as our targets:
PSR~J1823$-$3021A in NGC~6624, PSR~J1824$-$2452A in NGC~6626, PSR~J1835$-$3259B
in NGC~6652, and PSR~J1717$+$4308A in NGC~6341. Previous studies have suggested that
each of these MSPs may make a substantial contribution to the GeV $\gamma$-ray emission
of its host GC, and all four have sufficiently precise timing solutions for
reliable phase-folding analyses. The fundamental parameters of the selected MSPs and
their host GCs are summarized in Table~\ref{tab:msp_parameters}.

We analyzed $\sim$17.3~yr of Pass~8 data collected by \textit{Fermi}-LAT between
2008 August 4 and 2025 November 25 (MJD 54682.687--61004.000). We selected SOURCE-class
events including both front- and back-converting photons (\texttt{evclass=128} and
\texttt{evtype=3}) and used the \texttt{P8R3\_SOURCE\_V3} instrument response
functions. The analysis was performed over the energy range of 0.1--500.0~GeV.
For each target, events were extracted from a $20^\circ \times 20^\circ$ square
region of interest (RoI) centered on the MSP position. To suppress contamination
from Earth-limb $\gamma$-rays, we excluded events with zenith angles exceeding
$90^\circ$. We applied the standard data-quality selection,
\texttt{DATA\_QUAL > 0 \&\& LAT\_CONFIG == 1}, to exclude intervals with
non-nominal LAT operating conditions. All data reduction and likelihood analyses
were carried out using \texttt{Fermitools}-version of 2.0.19.

For each target, we constructed a source model for the whole likelihood analyses,
i.e. the phase-averaged spectral analyses,
including the cataloged point and extended $\gamma$-ray sources relevant to the
RoI, as well as the Galactic and isotropic diffuse backgrounds. The initial
source models were based on the fourth \textit{Fermi}-LAT source catalog Data
Release 4 \citep[4FGL-DR4;][]{bal+23},
and the corresponding XML files were generated with the \texttt{make4FGLxml.py}
script\footnote{\url{https://github.com/sfegan/glast-scripts/blob/main/make4FGLxml.py}}.
The diffuse background components were modeled using the Galactic interstellar
emission model \texttt{gll\_iem\_v07.fits} and the isotropic diffuse emission
model \texttt{iso\_P8R3\_SOURCE\_V3\_v1.txt}.

For each target, we allowed the normalizations and spectral parameters of
sources within $5^\circ$ of the target MSP to vary freely. For sources located
between $5^\circ$ and $10^\circ$ from the MSP, only the normalizations were
left free. In addition, we freed the normalizations of variable sources beyond
$10^\circ$ with \texttt{Variability\_Index > 72.44}. The normalizations of
the Galactic and isotropic diffuse components were also allowed to vary freely,
whereas all remaining source parameters were fixed at their 4FGL-DR4 catalog values.

We performed a standard binned-likelihood analysis of the four MSPs in the energy range of
$0.1$--$500.0$~GeV. The phase-averaged $\gamma$-ray spectra of MSPs are commonly described by
exponentially cut-off power-law models (PLEC), and their spectral energy distributions (SEDs) 
typically peak in the GeV band \citep{aaa+13}. Such spectral cutoffs are
broadly consistent with high-energy emission from relativistic electrons or positrons in pulsar
magnetospheres, including curvature-radiation scenarios \citep{hum05,smh+21}.
Following \citet{lsx+26}, we first fitted the phase-averaged $\gamma$-ray spectra using the
PLEC model, whose differential spectrum is given by
$\mathrm{d}N/\mathrm{d}E = N_0(E/E_0)^{-\gamma}\exp[-(E/E_{\rm c})^b]$, where $N_0$ is the
normalization, $E_0$ is the reference energy, $\gamma$ is the power-law index, $E_{\rm c}$ is
the cutoff energy, and $b$ controls the shape of the cutoff. In these initial fits, however,
the PLEC parameters were not well constrained. This result is likely related to the strong
correlations among the spectral parameters of the PLEC model (see Table~\ref{tab:spectral_results}), as discussed in the 4FGL-DR3
catalog \citep{aab+22}. Such correlations can become particularly important when the available
photon statistics are limited.

To mitigate these parameter correlations, we adopted the PLEC4 model (a reparameterized
form of the PLEC model), which has been used in Fermi-LAT pulsar catalog analyses.
In particular, its application to the Third
Fermi-LAT Catalog of Gamma-Ray Pulsars (3PC) was found to reduce parameter degeneracies and
improve fit convergence, thereby helping to constrain the spectral parameters of
pulsars \citep{smi+23}. PLEC4 is also widely used in the 4FGL-DR4 pulsar catalog \citep{aab+22,bal+23}.
These previous applications motivate its use as an alternative to the PLEC model in the
present analysis, although they do not imply that PLEC4 necessarily provides tighter
constraints for every individual source. The PLEC4 differential spectrum is given by
$\mathrm{d}N/\mathrm{d}E = N_0(E/E_0)^{\gamma_0+d/b} \exp\{[d/b^2][1-(E/E_0)^b]\}$,
where $N_0$ is the normalization, $E_0$ is the reference energy, $\gamma_0$ is the local
spectral index at $E_0$, and $d$ and $b$ characterize the spectral curvature and the shape
of the sub-exponential cutoff, respectively.
%Following \citet{aab+22}, the conventional cutoff energy can be obtained from the PLEC4 parameters as $E_{\rm c}=E_0(b^2/d)^{1/b}$. Thus, when $b$ is fixed, the cutoff energy is determined by the fitted curvature parameter $d$.

For all PLEC4 fits, we fixed the cutoff-shape parameter to $b=2/3$. This choice reduces
the number of free spectral parameters and provides a consistent parameterization for the
phase-averaged, off-pulse, and phase-resolved analyses. The fixed value of $b$ is a modeling
assumption adopted to improve parameter stability and maintain consistency among the different
analyses, it should not be interpreted as independent evidence for curvature-radiation emission.
The best-fit parameters of the PLEC4 and PLEC models for the four MSPs, obtained from the
phase-averaged spectral analyses, are summarized in Table~\ref{tab:spectral_results}.
We adopted the PLEC4 results as the reference spectral models for all subsequent analyses.

%For comparison, $b$ was left free in the PLEC fits, allowing the cutoff shape to be constrained directly by the data.

%\zmq{In Section 2.4, we use PLEC4 for the primary spectral-curvature analysis and additionally examine the conventional PLEC model for the two sources with reliable PLEC fits. In Section 2.5, we describe the parameter transformation used to derive the cutoff energy from the PLEC4 parameters.}

\begin{table*}
\centering
\caption{Best-fit Parameters from the Phase-averaged Analysis}
\label{tab:spectral_results}
\renewcommand{\arraystretch}{1.3} 
\begin{tabular}{l c c c @{\hspace{1.5em}} c c c c} 
\hline\hline
%\cmidrule(lr){2-8}
 & \multicolumn{3}{c}{PLEC4} & \multicolumn{4}{c}{PLEC} \\
\cmidrule(lr){2-4} \cmidrule(lr){5-8} 
\multicolumn{1}{c}{Pulsar name} & $\Gamma$ & $d$ & TS & $\Gamma$ & $b$ & $E_c$ (GeV) & TS \\
\cmidrule(lr){2-4} \cmidrule(lr){5-8} 
J1823$-$3021A & 1.85(12) & 0.43(1)  & 1206.2 & 0.77(2)  & 0.40(2)  & 0.10(1)  & 1196.9 \\ 
J1824$-$2452A & 2.14(1)  & 0.52(1)  & 1833.6 & 1.40(1)  & 1.28(2)  & 3.02(2)  & 1951.4 \\ 
J1835$-$3259B & 1.91(3)  & 0.66(4)  & 289.3  & 0.50(1)  & 0.46(1)  & 0.09(2)  & 282.9  \\ 
J1717$+$4308A & 1.89(25) & 0.65(27) & 97.3   & 0.92(38) & 0.86(12) & 0.55(17) & 97.8   \\  
\hline
\end{tabular}
\begin{flushleft} 
\small
\tablecomments{Results of the phase-averaged likelihood analysis for the four GC MSPs.
               All target MSPs were modeled with both the PLEC4 and PLEC spectral models. 
               In the PLEC4 model, $\Gamma$ and $d$ denote the photon index and exponential coefficient,
               respectively, and the cutoff-shape parameter $b$ was fixed at $2/3$ in all likelihood
               analyses. In the PLEC model, $\Gamma$ and $E_c$ denote the photon index and cutoff energy, respectively.
               Numbers in parentheses denote the $1\sigma$ uncertainties in the final
               quoted digits.}
\end{flushleft}
\end{table*}

\subsection{Ephemeris Update for PSR~J1824\texorpdfstring{$-$}{-}2452A}
\label{sec:uptime}

During the timing analysis described below, we found that the ephemerides provided by the Third \textit{Fermi}-LAT Catalog of Gamma-Ray Pulsars (3PC; \citealt{smi+23}) remained valid over the whole data span for PSR~J1823$-$3021A and PSR~J1835$-$3259B,
but not for PSR~J1824$-$2452A. Specifically, folding the LAT photons with the 3PC ephemeris for PSR~J1824$-$2452A produced a well-defined pulse profile only up to MJD~58939.896,
whereas the pulse signal became progressively smeared at later epochs, likely because
of accumulated phase drift associated with timing noise or inadequately modeled rotational
evolution. We therefore derived an updated timing solution for this pulsar.
First, we assigned rotational phases to photons collected between MJD~54682.683 and
58939.896 using the 3PC ephemeris and constructed an empirical Fourier template from
the resulting pulse profile. We then divided the whole LAT data set into consecutive
intervals of approximately 200~days and derived a pulse time of arrival (TOA) for each
interval by fitting the template to the photon phases using the unbinned maximum-likelihood
method described by \citet{ray+11}. The resulting TOAs were fitted iteratively with
\texttt{Tempo2} \citep{hem06}, with additional spin-frequency derivatives introduced as required to
account for the long-term rotational evolution and to extend the phase-connected solution
over the whole LAT observation span. The updated ephemeris provides a satisfactory
description of the TOAs over MJD~54682.687--61004.000. Its principal spin parameters are
$f=327.4055679765(13)\,\mathrm{Hz}$, $\dot{f}=-1.7347(8)\times10^{-13}\,\mathrm{s}^{-2}$,
and $\ddot{f}=-9.18(28)\times10^{-26}\,\mathrm{s}^{-3}$, corresponding to
the \texttt{Tempo2} parameters \texttt{F0}, \texttt{F1}, and \texttt{F2}, respectively;
the values in parentheses represent the $1\sigma$ uncertainties in the final quoted digits.

\subsection{Timing Analysis}
\label{sec:timing}

Using the best-fit models established in the whole likelihood analysis, we performed
a timing analysis to assign rotational phases to the \textit{Fermi}-LAT photons.
Reliable phase assignment is essential for the subsequent off-pulse and phase-resolved
spectral analyses, because the definitions of the off-pulse intervals and phase bins
are based directly on the $\gamma$-ray pulse profiles. To construct weighted
$\gamma$-ray pulse profiles, we selected photons within a circular aperture of
radius $3.16^\circ$ centered on each target MSP, following \citet{aaa+10}.
This aperture was chosen to retain adequate source-photon statistics while reducing
contamination from nearby sources and the two diffuse backgrounds. We then used
the \textit{Fermi}-tool \texttt{gtsrcprob}, together with the best-fit model,
to calculate the probability that each photon originated from the target MSP;
these probabilities were adopted as photon weights. For phase folding, we used the
3PC ephemerides for PSR~J1823$-$3021A and PSR~J1835$-$3259B \citep{smi+23},
the updated ephemeris derived in Section~\ref{sec:uptime} for PSR~J1824$-$2452A,
and the high-precision radio timing ephemeris for PSR~J1717$+$4308A obtained from
the Five-hundred-meter Aperture Spherical radio Telescope (FAST) observations in
\citet{prl+20,pqm+21}. The weighted photons were coherently folded with
\texttt{Tempo2} and its \textit{Fermi} plug-in \citep{hem06}, which transforms the photon arrival
times to the Solar System barycenter before evaluating the rotational phase.

\begin{figure*}[htbp]
    \centering
    \includegraphics[width=0.49\textwidth]{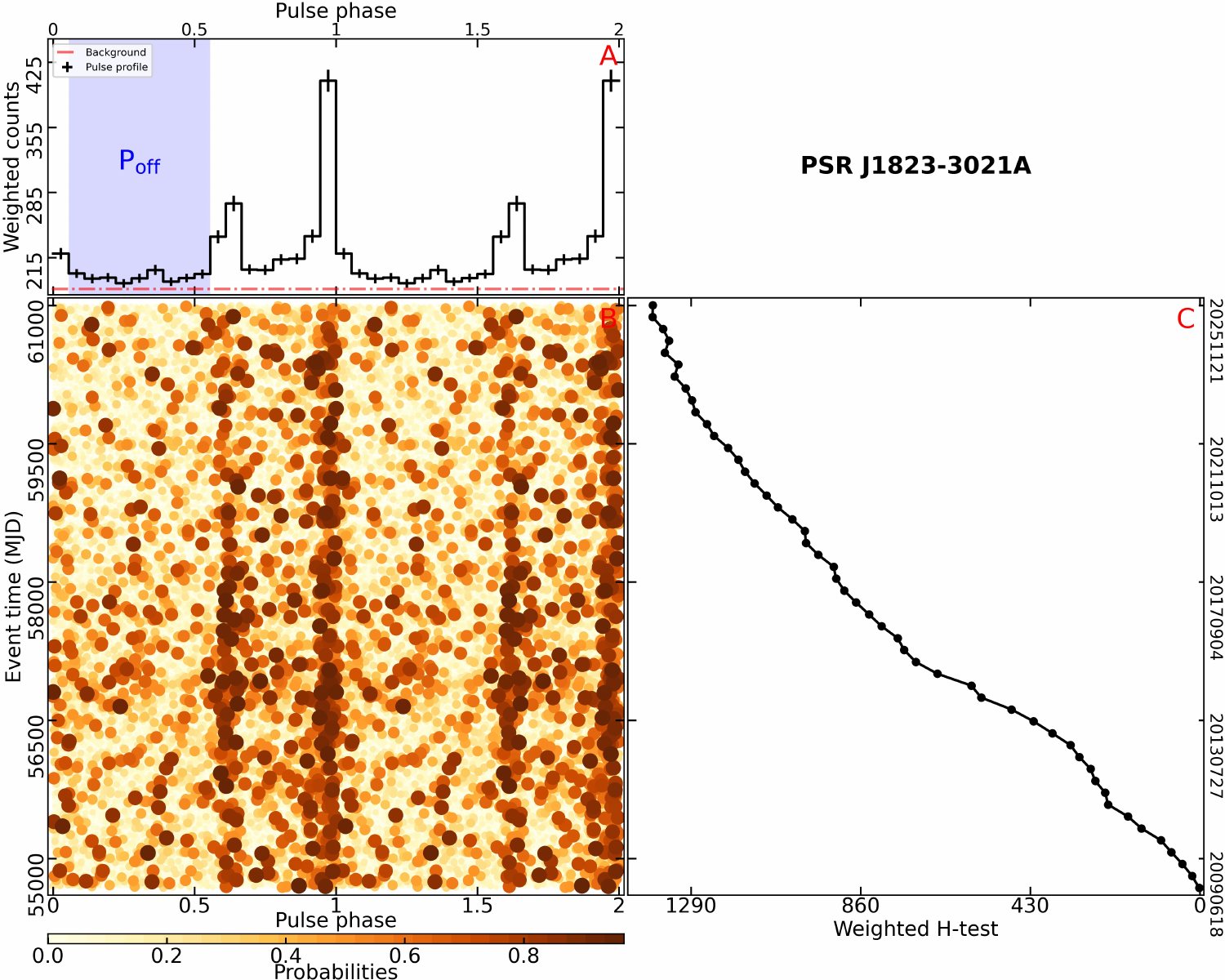}
    \includegraphics[width=0.49\textwidth]{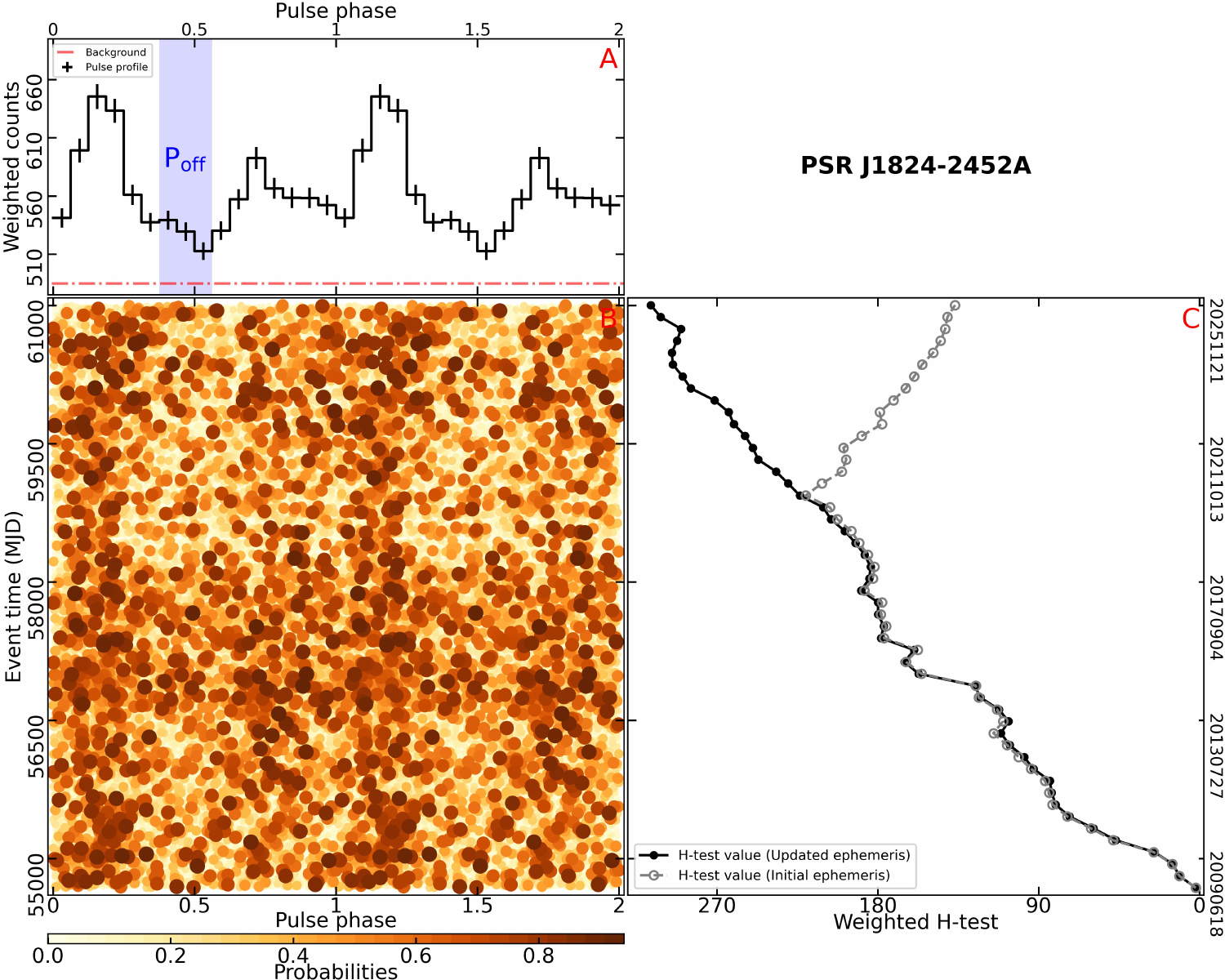}
    \includegraphics[width=0.49\textwidth]{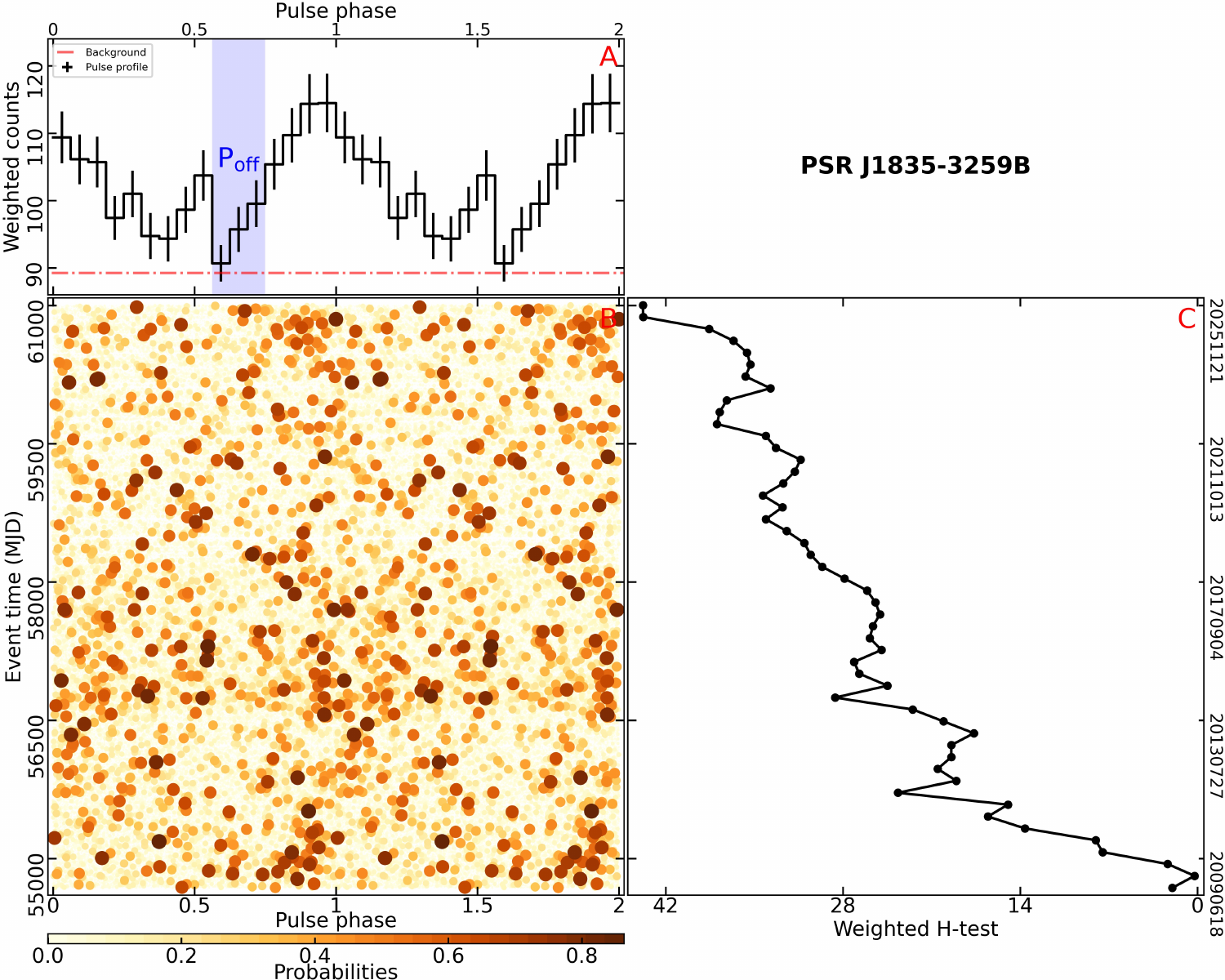}
    \includegraphics[width=0.49\textwidth]{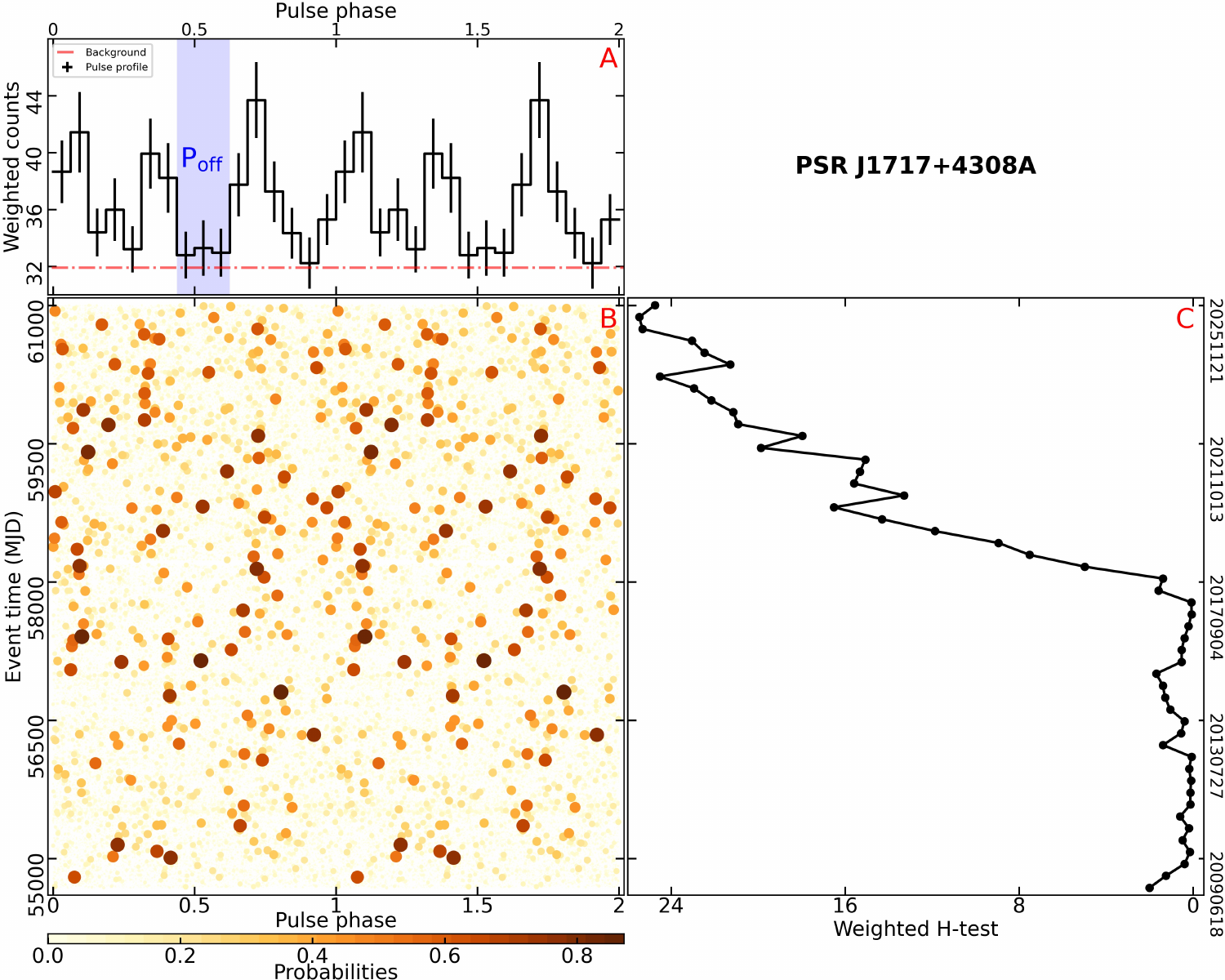}
    \caption{Timing-analysis results for the four GC MSPs. The four main panels correspond
             to PSR~J1823$-$3021A (upper left), PSR~J1824$-$2452A (upper right),
             PSR~J1835$-$3259B (lower left), and PSR~J1717$+$4308A (lower right).
             Within each main panel, subpanels (A)--(C) show, respectively: (A) the integrated,
             probability-weighted $\gamma$-ray pulse profile, displayed over two rotational
             cycles for clarity, with the horizontal dash-dotted line denoting the estimated
             background level; (B) the two-dimensional phaseogram, in which photon rotational
             phase is shown as a function of time and the color scale indicates the photon
             probability weights; and (C) the cumulative weighted H-test statistic as
             a function of time over the full \textit{Fermi}-LAT observation span.
             For PSR~J1824$-$2452A, the H-test results obtained using the original 3PC ephemeris
             are additionally shown as gray circles in subpanel (C) of the upper-right main panel.}
    \label{fig:timing}
\end{figure*}

The weighted $\gamma$-ray pulse profiles for the target MSPs are presented in
panel A of Figure~\ref{fig:timing}, where the weighted counts in each phase bin and
their uncertainties were calculated following \citet{aaa+13}. Panel B shows
the corresponding two-dimensional phaseograms, with photon rotational phase plotted
as a function of time. We also calculated the weighted H-test statistic
following \citet{ker11}, using the \texttt{gtsrcprob}-derived source probabilities
as photon weights; panel C presents the cumulative H-test statistic as a function of
time over the full \textit{Fermi}-LAT observation span. As shown in Figure~\ref{fig:timing},
all four target MSPs exhibit prominent $\gamma$-ray pulsations. The phaseograms show
that the pulse profiles remain phase coherent throughout the 17.3~yr data set,
supporting the validity of the timing ephemerides adopted in this work. In addition,
the cumulative H-test statistics increase steadily with time, demonstrating
the persistent detection of pulsed $\gamma$-ray emission from all four MSPs.
These well-defined and temporally stable pulse profiles provide the basis for
the determination of off-pulse phase intervals in the following section.

For comparison, we also show the H-test results obtained using the original,
uncorrected 3PC ephemeris for PSR~J1824$-$2452A as a gray dashed line in the top right
of Figure~\ref{fig:timing}; the marked degradation in pulsation significance after
MJD~58939.896 clearly demonstrates that this ephemeris is no longer valid at later epochs.

\subsection{Off-pulse Emission Analysis}
\label{sec:off_pulse}

To determine whether the four target MSPs exhibit significant $\gamma$-ray emission outside
their main pulsed components, we performed dedicated off-pulse analyses. Based on the
probability-weighted $\gamma$-ray pulse profiles, we defined the off-pulse interval for each
MSP as the phase range away from the main pulse peaks in which the profile is consistent with
the background level. The adopted off-pulse phase ranges, listed in the second column of
Table~\ref{tab:offpulse_results}, are consistent with those reported
in \citet{fre+11, whw+13, joh+13, zxw22, zxw+23}. We then selected \textit{Fermi}-LAT photons
within the corresponding off-pulse intervals and performed a binned maximum-likelihood analysis
to test for significant off-pulse $\gamma$-ray emission from each target MSP.

In the off-pulse likelihood analysis, the normalization and spectral parameters of
each target MSP were allowed to vary freely, as were the normalizations of the Galactic
and isotropic diffuse emission components; all parameters of the remaining sources
were fixed at their best-fit values obtained from the phase-averaged analysis.
To account for the reduced effective exposure resulting from the phase selection,
the exposure of the phase-filtered data, or equivalently the normalizations of the
fixed model components, was scaled by the fractional width of the corresponding
off-pulse phase interval. This procedure ensures that the model-predicted photon
counts are consistent with the phase-filtered data. We quantified the significance
of off-pulse emission using the likelihood TS. We adopted $\mathrm{TS}>16$,
corresponding approximately to a $4\sigma$ detection, as the criterion for
significant off-pulse $\gamma$-ray emission from a target MSP.

The best-fit spectral results for the off-pulse intervals are summarized in
Table~\ref{tab:offpulse_results}. According to the adopted criterion of $\mathrm{TS}>16$,
significant off-pulse $\gamma$-ray emission is detected from PSR~J1823$-$3021A,
PSR~J1824$-$2452A, and PSR~J1835$-$3259B, with $\mathrm{TS}$ values of 61.48, 79.24,
and 16.70, respectively. In contrast, PSR~J1717$+$4308A is not significantly detected
in its off-pulse interval, with $\mathrm{TS}=14.43$, below the adopted threshold.
Consequently, PSR~J1717$+$4308A does not meet the selection criterion for the subsequent
phase-resolved analysis and is not considered further in the remainder of this work. 
As an additional check, we also evaluated the off-pulse emission using the conventional PLEC model
for the two MSPs for which the PLEC fits converged reliably. The PLEC analyses were performed using the same data selection and analysis settings as those adopted for the PLEC4 analyses. 
The resulting TS values are 16.39 for PSR~J1835$-$3259B and 14.80 for PSR~J1717$+$4308A, respectively. 
These results yield the same off-pulse detection classifications as those obtained with the PLEC4 model, indicating that the off-pulse source-selection results are consistent between the two spectral models.

\begin{table*}
\centering
\caption{Phase-resolved spectral results of the off-pulse intervals}
\renewcommand{\arraystretch}{1.2}
\begin{tabular}{lccccc}
\hline\hline
 & Off-pulse & \multicolumn{4}{c}{Parameter Values} \\
\cline{3-6}
\multicolumn{1}{c}{Pulsar name} & Off-pulse Interval & $\Gamma$ & $d$ & TS & $\text{TS}_{\text{cutoff}}$ \\
\hline
J1823$-$3021A & 0.0556--0.5556 & 2.18(22) & 0.87(62) & 61.48 & 14.80 \\
J1824$-$2452A & 0.375--0.5625 & 2.32(9)  & 0.49(21) & 79.24 & 12.38 \\
J1835$-$3259B & 0.5625--0.75   & 2.22(19) & 1.06(60) & 16.70 & 8.08 \\
J1717$+$4308A & 0.4375--0.625  & 2.22(3)  & 0.62(22) & 14.43 & -- \\
\hline
\end{tabular}
\begin{flushleft} 
\small
\tablecomments{Results of the off-pulse likelihood analysis for the four GC MSPs. The adopted
               off-pulse phase intervals are consistent with those reported
               in \citet{fre+11, whw+13, joh+13, zxw22, zxw+23}. The definitions of the PLEC4
               spectral parameters are the same as those given in Table~\ref{tab:spectral_results}.
               The $\mathrm{TS}$ values were obtained from likelihood fits to the data selected
               within the corresponding off-pulse intervals using the PLEC4 model. According to
               the adopted criterion of $\mathrm{TS}>16$, PSRs~J1823$-$3021A, J1824$-$2452A, and
               J1835$-$3259B exhibit significant off-pulse $\gamma$-ray emission. The statistic
               $\mathrm{TS}_{\rm cutoff}^{\rm PLEC4}$ is the likelihood-ratio TS quantifying the preference
               for the PLEC4 model over a PL model. A dash (--) indicates that
               $\mathrm{TS}_{\rm cutoff}^{\rm PLEC4}$ was not computed for PSR~J1717$+$4308A because
               its off-pulse emission did not satisfy the adopted detection criterion.}
\end{flushleft}
\label{tab:offpulse_results}
\end{table*}

A comparison of the spectral parameters derived from the phase-averaged (Table~\ref{tab:spectral_results})
and off-pulse (Table~\ref{tab:offpulse_results}) analyses provides useful insight into
the origin of the $\gamma$-ray emission. For the three MSPs with significant off-pulse
detections, the off-pulse photon indices ($\Gamma \simeq 2.18$--$2.32$) are systematically
softer than their phase-averaged values ($\Gamma \simeq 1.85$--$2.14$). 
Although these softer spectra could suggest a contribution from an unpulsed component,
such as PWN emission or the integrated emission of unresolved MSPs in
the host GCs, we further examined the off-pulse spectral shapes by
comparing the PLEC4 and PL models, as discussed below.
This result suggests that
the off-pulse emission retains a curved spectral character broadly compatible with
magnetospheric MSP emission; however, spectral curvature alone does not exclude a contribution
from unresolved cluster MSPs, whose collective spectrum may also exhibit a cutoff.
Similar off-pulse spectral cutoffs have been interpreted as evidence favoring
a magnetospheric origin in field MSPs, but dedicated spatial and spectral tests are
required to distinguish this possibility from alternative components in GC environments \citep{lsx+26}.

To further investigate the spectral origin of the off-pulse emission, we compared
a simple power-law (PL) model with the PLEC4 for the three MSPs with significant off-pulse detections.
For each MSP, we performed independent maximum-likelihood fits to the off-pulse data
using the PL and PLEC4 models. Although PLEC4 was adopted as the baseline model in
the phase-averaged analysis, neither model was assumed \textit{a priori} to describe
the off-pulse spectrum; instead, they were treated as competing spectral hypotheses.
The preference for a cutoff spectrum was quantified using the likelihood-ratio TS,
$\mathrm{TS}_{\rm cutoff}^{\rm PLEC4} =2\ln(\mathcal{L}_{\rm PLEC4}/\mathcal{L}_{\rm PL})$,
where $\mathcal{L}_{\rm PL}$ and $\mathcal{L}_{\rm PLEC4}$ are the maximum likelihoods
obtained with the PL and PLEC4 models, respectively \citep{aje+10}.
We adopted $\mathrm{TS}_{\rm cutoff}^{\rm PLEC4}>9$
as evidence for a statistically significant preference for the PLEC4 model over the PL model,
corresponding approximately to a $3\sigma$ significance level.

\begin{figure*}[htbp]
    \centering
    \includegraphics[width=0.44\textwidth]{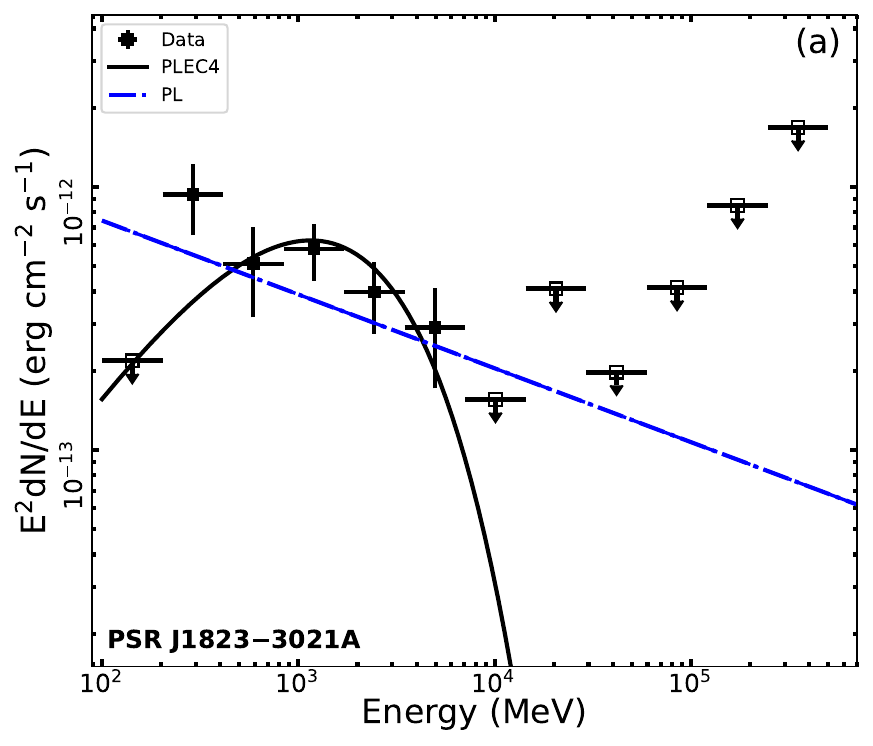}
    \includegraphics[width=0.44\textwidth]{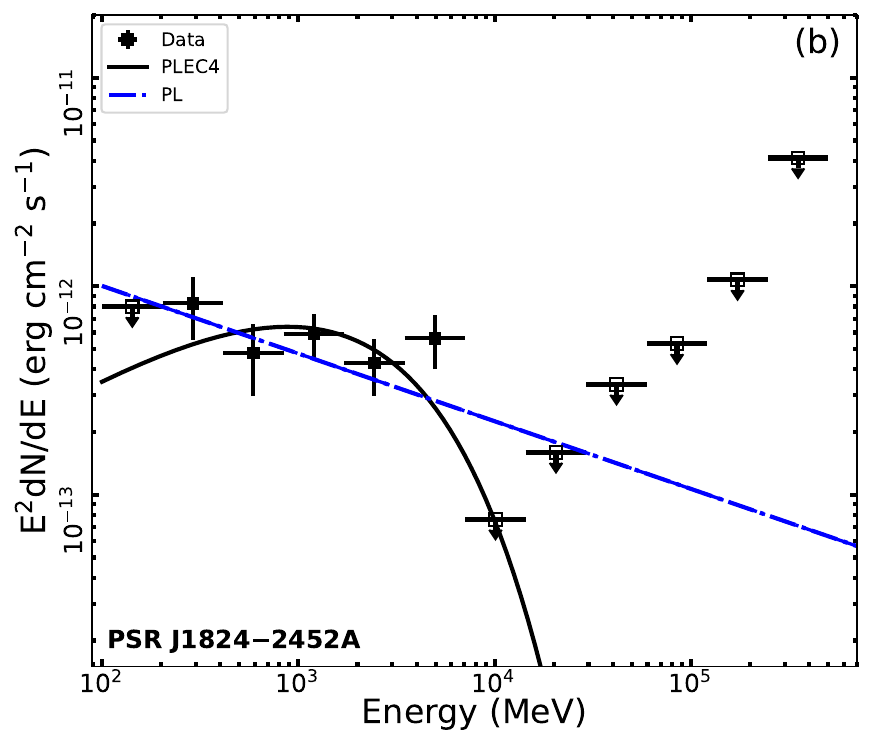}
    \includegraphics[width=0.44\textwidth]{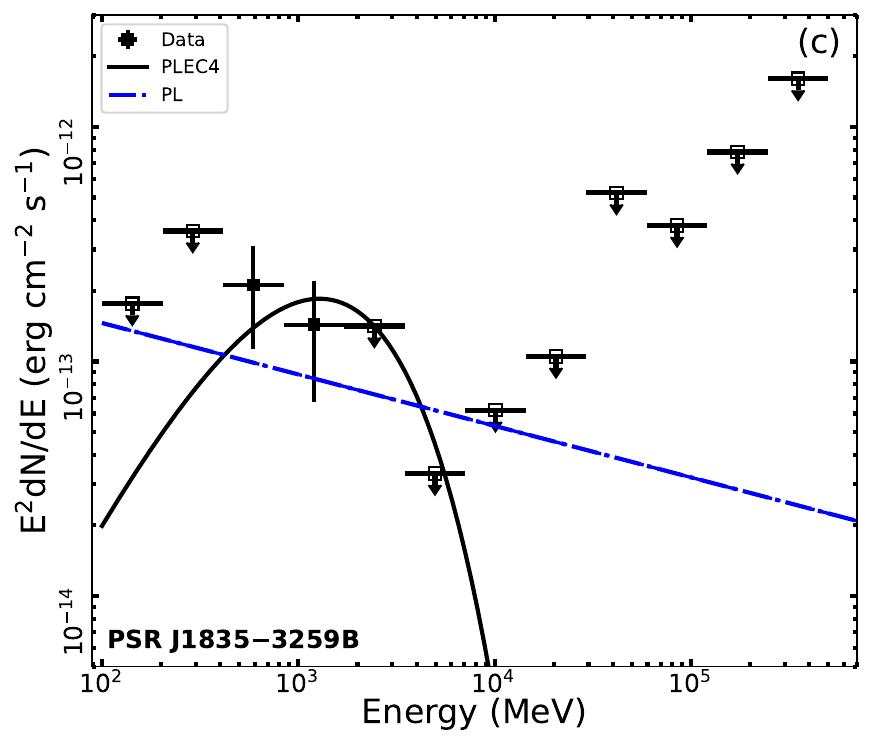}
    \caption{SEDs of the off-pulse emission from the three GC MSPs with significant
             off-pulse detections. From top to bottom, panels (a)--(c) correspond to
             PSR~J1823$-$3021A, PSR~J1824$-$2452A, and PSR~J1835$-$3259B, respectively.
             The best-fit PLEC4 and PL models are shown as black solid and blue dashed
             curves, respectively. Data points denote the best-fit energy fluxes
             in individual energy bins with $\mathrm{TS}>4$, whereas downward arrows
             indicate the 95\% upper limits for bins with $\mathrm{TS}<4$.}
    \label{fig:offpulse_sed}
\end{figure*}

The off-pulse spectra of PSR~J1823$-$3021A and PSR~J1824$-$2452A show statistically
significant preferences for the PLEC4 model over a simple PL model, with $\mathrm{TS}_{\rm cutoff}^{\rm PLEC4}=14.80$ and 12.38, respectively, both exceeding the adopted threshold of 9. These results
indicate significant spectral curvature in the off-pulse emission of both MSPs,
at a significance $>3\sigma$. The presence of a GeV spectral cutoff is consistent with a predominantly magnetospheric
origin for their off-pulse $\gamma$-ray emission, although this feature alone is insufficient
to establish curvature radiation as the underlying emission mechanism. At the same time,
this result disfavors an origin dominated by a typical PWN-like component with a
featureless PL spectrum. However, the spectral cutoff
alone cannot exclude a contribution from unresolved MSPs in the host GC, whose integrated
emission may also exhibit a cutoff spectrum. In contrast, PSR~J1835$-$3259B has
$\mathrm{TS}_{\rm cutoff}^{\rm PLEC4}=8.08$, which does not reach the adopted threshold for
a significant preference for the PLEC4 model. Given its marginal off-pulse detection
($\mathrm{TS}=16.70$), the available photon statistics are insufficient to distinguish
robustly between the PL and PLEC4 models.
As an additional check, we performed a spectral-curvature test using the conventional PLEC model 
for PSR~J1835$-$3259B, the only one of the three MSPs with significant off-pulse $\gamma$-ray emission for which a reliable PLEC fit was obtained. We quantified the preference for the PLEC model over the PL model using the likelihood-ratio test statistic,
$\mathrm{TS}_{\rm cutoff}^{\rm PLEC}=2\ln(\mathcal{L}_{\rm PLEC}/\mathcal{L}_{\rm PL})$,
where $\mathcal{L}_{\rm PLEC}$ and $\mathcal{L}_{\rm PL}$ are the maximum likelihoods obtained with the PLEC and PL models, respectively. The resulting $\mathrm{TS}_{\rm cutoff}^{\rm PLEC}$ is 7.90, below the adopted threshold of 9. This result is consistent with that obtained using PLEC4,
for which $\mathrm{TS}_{\rm cutoff}^{\rm PLEC4}=8.08$, as neither model shows a statistically significant preference over the PL model for this source. We do not report PLEC-based curvature-test results for the other two MSPs with significant off-pulse emission because reliable PLEC fits could not be obtained for them. To illustrate the off-pulse spectral fits,
we constructed spectral energy distributions (SEDs) over 0.1--500.0~GeV by dividing
the data into 12 logarithmically spaced energy bins and performing an independent
maximum-likelihood fit in each bin. In these fits, the target MSP was modeled with
a PL spectrum with its photon index fixed at 2. For bins with $\mathrm{TS}\geq4$,
we report the best-fit fluxes and their statistical uncertainties;
for bins with $\mathrm{TS}<4$, we derive 95\% upper limits. The resulting off-pulse SEDs
are shown in Figure~\ref{fig:offpulse_sed}.

\subsection{Phase-resolved Analysis}
\label{sec:phase_resolved_analysis}

Based on the off-pulse analysis presented in Section~\ref{sec:off_pulse}, significant
off-pulse $\gamma$-ray emission was detected from PSR~J1823$-$3021A, PSR~J1824$-$2452A,
and PSR~J1835$-$3259B, whereas PSR~J1717$+$4308A did not satisfy the adopted detection
criterion and was therefore excluded from further analysis. The significant off-pulse
emission and the spectral curvature detected in some of these sources motivate a more
detailed investigation of the evolution of their spectral properties with rotational
phase. We therefore performed a phase-resolved spectral analysis for the three selected
MSPs. Following the phase-binning scheme established from the weighted $\gamma$-ray
pulse profiles, we divided one complete rotational cycle into multiple phase bins and
modeled the spectrum independently in each bin.

\begin{figure*}[htbp]
    \centering
    \includegraphics[width=0.49\textwidth]{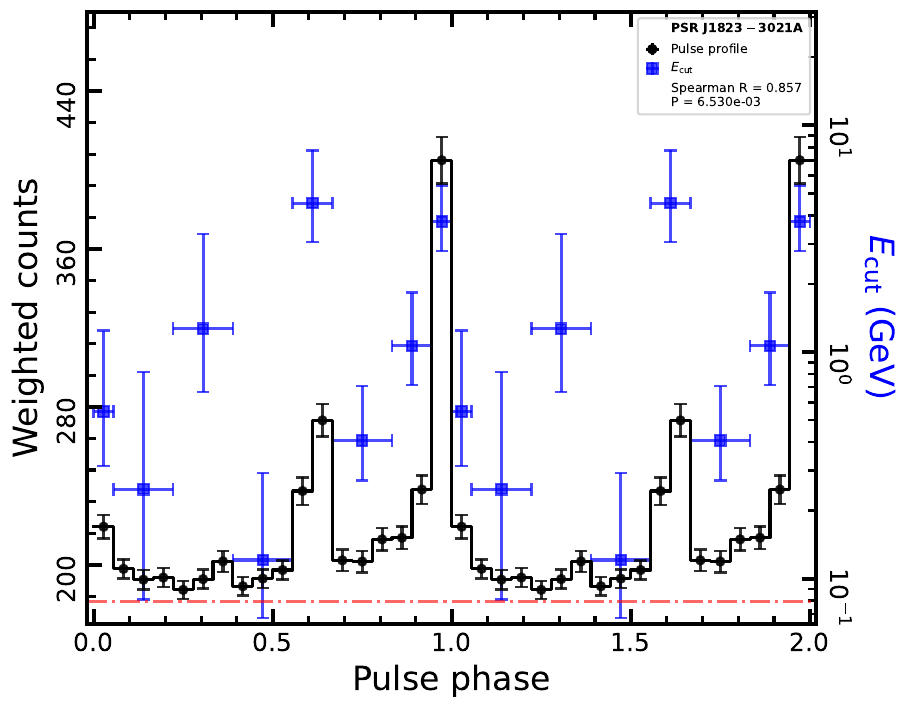}
    \includegraphics[width=0.49\textwidth]{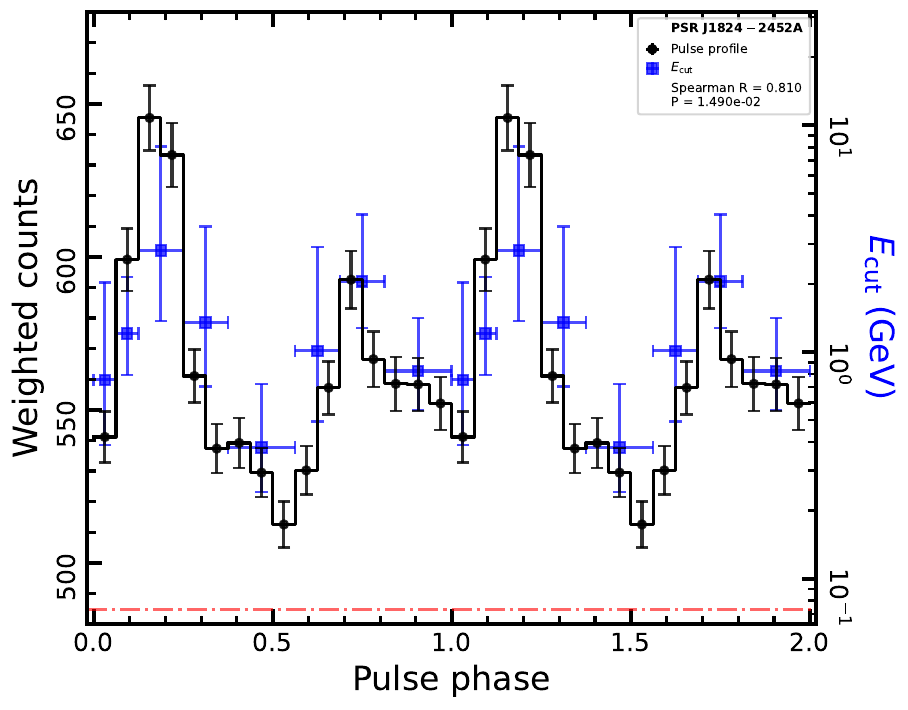}
    \includegraphics[width=0.49\textwidth]{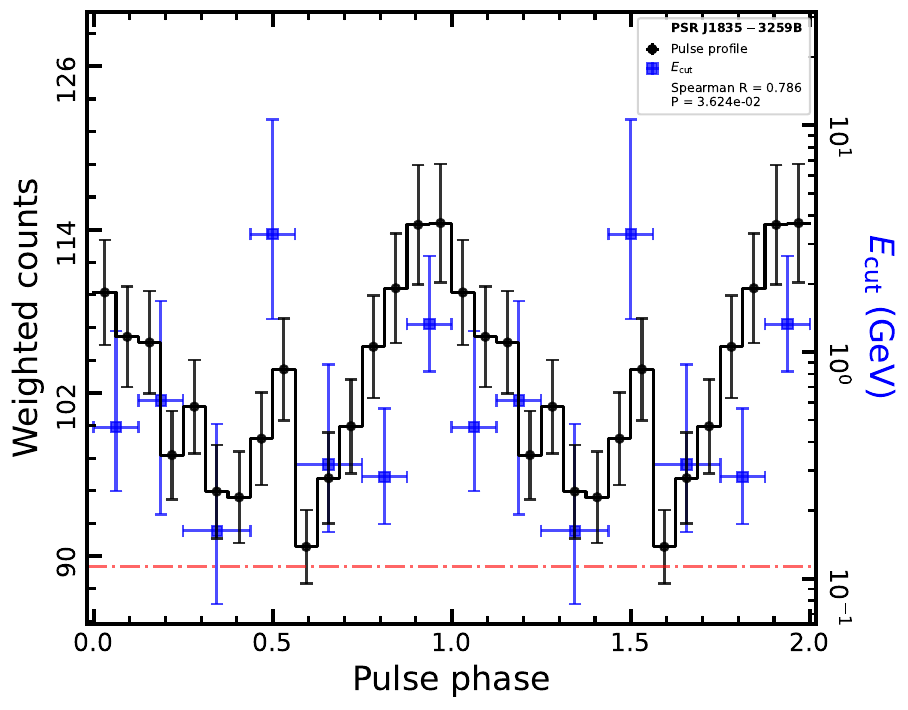}
    \caption{Phase-resolved spectral results for the three GC MSPs with significant
             off-pulse emission. Each panel shows the probability-weighted $\gamma$-ray
             pulse profile and the evolution of the $E_{\rm cut}$, over the rotational
             cycle. The black histograms, referenced to the left-hand axis,
             show the probability-weighted $\gamma$-ray pulse profiles, identical to
             those presented in Figure~\ref{fig:timing}. The blue points, referenced to
             the right-hand axis, represent the best-fit values of $E_{\rm cut}$ in the
             individual phase bins; their error bars indicate the corresponding
             statistical uncertainties. The red horizontal dash-dotted line denotes
             the estimated background level of the pulse profile.}
    \label{fig:phase_resolved_results}
\end{figure*}

For the phase-resolved spectral fits, we adopted the same PLEC4 spectral model used in
the off-pulse analysis. Following the baseline modeling scheme described in
Section~\ref{sec:off_pulse}, we assumed a common photon index, $\Gamma$, across all
phase bins and fixed it to the best-fit value obtained from the phase-averaged analysis.
The spectral evolution across the rotational phase was therefore characterized primarily
by variations in the cutoff energy and flux normalization. Because the cutoff parameter
of the PLEC4 model is not directly equivalent to the cutoff energy of the conventional
exponentially cutoff power-law model, we converted the fitted PLEC4 parameters into
an effective $E_{\rm cut}$. Following the prescription given in
the 3PC \citep{smi+23}, we calculated $E_{\rm cut}=E_0\left(b^2/d\right)^{1/b}$, where
$E_0$ is the scale energy, $b$ is the cutoff-shape parameter fixed at $2/3$
in our analysis, and $d$ is the exponential coefficient of the PLEC4 model.
This relation provides an effective cutoff energy corresponding
to the conventional exponentially cutoff power-law parameterization,
without requiring an independent fit with the conventional PLEC model. 
We applied this conversion to the best-fit PLEC4 parameters in 
each phase bin.
The statistical uncertainties of $E_{\rm cut}$ were propagated from the uncertainties
of the fitted parameter $d$ using a Monte Carlo (MC) procedure.
The phase-resolved spectral results are presented in Figure~\ref{fig:phase_resolved_results}.
In each panel, the black histogram shows the probability-weighted $\gamma$-ray pulse
profile, the blue points represent the derived $E_{\rm cut}$, in the individual phase
bins, and the red dash-dotted line denotes the estimated background level.
For all three MSPs, $E_{\rm cut}$ exhibits clear phase-dependent modulation and generally
follows the morphology of the pulse profile, reaching higher values near the main pulse
peaks and decreasing toward the inter-pulse and off-pulse regions. We quantified this
trend using a Spearman rank-correlation test between the phase-resolved energy flux and
$E_{\rm cut}$. A statistically significant positive correlation is found for each of the
three MSPs ($p<0.05$), indicating that the brighter phase intervals tend to have higher
cutoff energies. This behavior is consistent with previous phase-resolved studies of
bright $\gamma$-ray pulsars, including Vela, the Crab, and Geminga, where the spectral
cutoff was generally found to become harder near the peaks of the $\gamma$-ray light
curves \citep{dec13}.

\subsection{Phase-resolved \texorpdfstring{$L_{\gamma}$--$E_{\rm cut}$}{Lg--Ecut} Relation}
\label{sec:LandEc}

The phase-resolved results for the three selected MSPs show that both the energy flux
and $E_{\rm cut}$ vary systematically with rotational phase and generally follow the
morphology of the $\gamma$-ray pulse profiles. The main pulse intervals account for
most of the observed high-energy emission and tend to exhibit higher cutoff energies,
whereas the inter-pulse and off-pulse intervals show weaker emission with lower
$E_{\rm cut}$ values. Together with the significant off-pulse detections reported in Section~\ref{sec:off_pulse}, these phase-dependent spectral properties support the
interpretation that the off-pulse emission is physically connected to
the pulsed magnetospheric emission. For PSR~J1823$-$3021A and PSR~J1824$-$2452A, the
statistically significant preference for a cutoff spectrum in the off-pulse interval
further supports a predominantly magnetospheric origin. Although the off-pulse
spectrum of PSR~J1835$-$3259B does not show a statistically significant preference
for the PLEC4 model over a simple power law, its significant off-pulse detection and
phase-dependent spectral behavior are consistent with the presence of
a broad magnetospheric emission component. These results disfavor an interpretation
in which the detected off-pulse emission is dominated solely by background-model
residuals or a steady unresolved cluster component. The implications for the origin
of the off-pulse emission are discussed further in Section~\ref{sec:summary_discussion}.

Because the observed energy flux depends on the source distance, we quantified the
relation between the phase-resolved $E_{\rm cut}$, and the $\gamma$-ray luminosity ($L_\gamma$)
to investigate the physical connection between spectral shape and emission intensity
across the rotational cycle. We define the phase-resolved
isotropic $\gamma$-ray luminosity as $L_{\gamma}=4\pi D^{2}G$, where $D$ is the
distance to the MSP, as listed in Table~\ref{tab:msp_parameters}, and $G$ is the
integrated energy flux in the corresponding phase bin over the energy range adopted
in this work. This definition is equivalent to setting the beaming correction factor
to $f_{\Omega}=1$. Although the true luminosity
is generally $L_{\gamma}=4\pi f_{\Omega}D^{2}G$, a phase-dependent determination of
$f_{\Omega}$ requires detailed modeling of the pulsar emission geometry and is beyond
the scope of this work.

Using the phase-resolved measurements described above, we investigated the correlation
between $\log_{10}(L_{\gamma})$ and $\log_{10}(E_{\rm cut})$. We modeled the relation
as a linear trend in logarithmic space with the \texttt{LinearRegressionwithErrors}
function implemented in the \texttt{astroML}\footnote{\url{https://github.com/astroML/astroML}} Python package, which follows
the Bayesian linear-regression method of \citet{kel07}. This method incorporates
measurement uncertainties in both $\log_{10}(L_{\gamma})$ and $\log_{10}(E_{\rm cut})$,
thereby providing a statistically consistent estimate of the correlation slope
and its uncertainty.

\begin{figure}[htbp]
    \centering
    \includegraphics[width=0.45\textwidth]{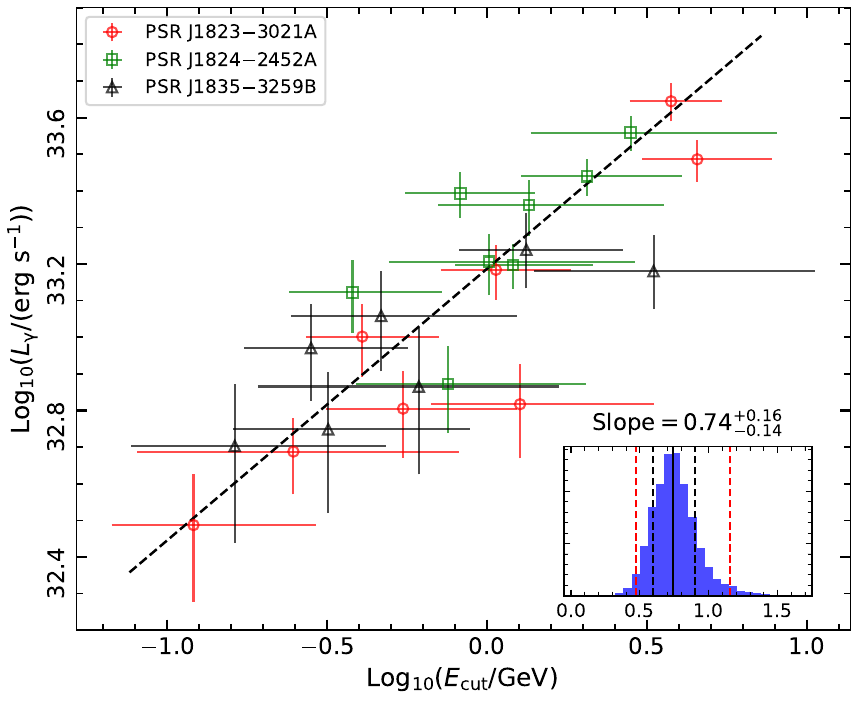}
    \caption{Bayesian linear regression between $\log_{10}(L_{\gamma})$ and
             $\log_{10}(E_{\rm cut})$, using the phase-resolved spectral results from the
             three GC MSPs with significant off-pulse emission. Red circles, green squares,
             and black triangles represent the phase-bin measurements of PSR~J1823$-$3021A, PSR~J1824$-$2452A, and PSR~J1835$-$3259B, respectively. The black dashed line
             shows the best-fit relation, $L_{\gamma}\propto E_{\rm cut}^{\alpha}$, with
             $\alpha=0.74^{+0.16}_{-0.14}$ (68\% credible interval). The inset presents
             the posterior probability distribution of $\alpha$; the black solid line marks
             the posterior median, while the dashed black and red lines indicate the 68\%
             and 95\% credible intervals, respectively.}
    \label{fig:logL_logEcut}
\end{figure}

The results of the phase-resolved correlation analysis are presented in
Figure~\ref{fig:logL_logEcut}. We find a statistically significant positive correlation
between $\log_{10}(L_{\gamma})$ and $\log_{10}(E_{\rm cut})$,
described by $L_{\gamma}\propto E_{\rm cut}^{\alpha}$ with a best-fit slope
of $\alpha=0.74^{+0.16}_{-0.14}$. This result demonstrates that the phase-resolved
$\gamma$-ray luminosity and spectral cutoff energy are positively coupled, although the
relatively shallow slope indicates that the $L_{\gamma}$ increases more gradually than
the $E_{\rm cut}$. Our measured slope is substantially smaller than the value
of $\alpha=2.31^{+0.22}_{-0.25}$ reported by \citet{lsx+26}, which is consistent with
the $\alpha=16/7\simeq2.29$ scaling expected in the ECS curvature-radiation framework \citep{khk+19}.
We note that \citet{lsx+26} used the $\gamma$-ray pseudoluminosity, whereas our analysis uses the isotropic $\gamma$-ray luminosity.
The difference suggests that the phase-resolved $L_{\gamma}$--$E_{\rm cut}$ relation
in GC MSPs may not be described by the same simple scaling inferred for field MSPs.
The origin of this discrepancy is likely complex and multi-faceted,
involving a combination of observational, environmental modulation,
and physical effects that are particularly relevant in the globular
cluster context. We will further explore these findings and
possibilities in Section~\ref{sec:summary_discussion}.

%Possible causes include differences in magnetospheric geometry and beaming, contamination from unresolved cluster emission, and limitations associated with the adopted isotropic luminosity approximation; these possibilities are discussed further in Section~\ref{sec:summary_discussion}.

%The difference suggests that the phase-resolved $L_{\gamma}$--$E_{\rm cut}$ relation in GC MSPs may not be described by the same simple scaling inferred for field MSPs. Possible causes include differences in magnetospheric geometry and beaming, contamination from unresolved cluster emission, and limitations associated with the adopted isotropic luminosity approximation; these possibilities are discussed further in Section~\ref{sec:summary_discussion}.
%
%This discrepancy may suggest that environmental factors—such as the intense background radiation fields within globular clusters or interactions between the pulsar wind and binary companion winds—regulate the magnetospheric acceleration process, or that additional non-magnetospheric emission components contaminate the phase-resolved intervals.We will further discuss these findings and their relationship with the MSP evolutionary environment in Section~\ref{sec:summary_discussion}.

\section{SUMMARY and DISCUSSION}
\label{sec:summary_discussion}

In this work, we analyzed $\sim$17.3~yr of \textit{Fermi}-LAT Pass~8 data between 0.1
and 500.0~GeV for four MSPs hosted by GCs. Significant off-pulse $\gamma$-ray emission,
defined by the criterion $\mathrm{TS}>16$, was detected from three sources:
PSR~J1823$-$3021A, PSR~J1824$-$2452A, and PSR~J1835$-$3259B. For PSR~J1823$-$3021A and
PSR~J1824$-$2452A, the off-pulse spectra show statistically significant preferences
for the PLEC4 model over a simple PL model, with $\mathrm{TS}_{\rm cutoff}^{\rm PLEC4}=14.80$ and
12.38, respectively. The resulting GeV spectral curvature supports a predominantly
magnetospheric origin for their off-pulse emission, although it does not by itself
exclude a contribution from unresolved MSPs in the host GCs. PSR~J1835$-$3259B has a
marginal off-pulse detection, with $\mathrm{TS}=16.70$, and its spectrum does not
show a statistically significant preference for the PLEC4 model
($\mathrm{TS}_{\rm cutoff}^{\rm PLEC4}=8.08$); the available photon statistics are therefore
insufficient to distinguish robustly between the PL and PLEC4 spectral descriptions.

We performed phase-resolved spectral analyses for the three MSPs with significant
off-pulse detections. As shown in Figure~\ref{fig:phase_resolved_results},
all three sources exhibit phase-dependent variations in the $E_{\rm cut}$
that generally follow the morphology of their weighted $\gamma$-ray
pulse profiles. For PSR~J1823$-$3021A, which shows a narrow main pulse, $E_{\rm cut}$
reaches its highest value near the pulse peak and decreases in phase intervals away
from the peak. PSR~J1824$-$2452A exhibits a more complex multi-peaked pulse profile,
and its phase-resolved $E_{\rm cut}$ shows local enhancements broadly associated with
the principal pulse components, with lower values in the inter-pulse intervals.
Despite the more limited photon statistics of PSR~J1835$-$3259B, its $E_{\rm cut}$
values also tend to be higher in the pulse-dominated phase intervals and lower toward
the off-pulse region. The positive correlations between phase-resolved energy flux
and $E_{\rm cut}$ found for all three MSPs indicate that brighter phase intervals
generally have harder cutoff spectra, supporting a physical connection between
emission intensity and spectral shape across the rotational cycle.

Taken together, the phase-resolved results for the three MSPs indicate that both
the $\gamma$-ray energy flux and $E_{\rm cut}$ vary systematically with rotational
phase and generally follow the morphology of the weighted pulse profiles.
The pulse-dominated phase intervals produce the strongest emission and tend to
exhibit higher cutoff energies, whereas the off-pulse intervals retain weaker but
significant $\gamma$-ray emission. This phase-dependent behavior supports
a physical connection between the off-pulse and pulsed emission components.
For PSR~J1823$-$3021A and PSR~J1824$-$2452A, the statistically significant
preference for a cutoff spectrum in the off-pulse interval provides additional
support for a predominantly magnetospheric origin. Although the off-pulse spectrum
of PSR~J1835$-$3259B does not significantly favor the PLEC4 model over a simple
power law, its phase-resolved behavior remains consistent with
a broad magnetospheric emission component. Therefore, our results disfavor
an interpretation in which the detected off-pulse emission is dominated solely
by background-model residuals or a steady unresolved cluster component,
while a contribution from unresolved MSPs in the host GCs cannot be excluded.
Similar phase-resolved correlations between emission intensity and $E_{\rm cut}$
have been interpreted as evidence for magnetospheric emission across
a broad range of rotational phases in field MSPs \citep{lsx+26}.

To investigate the connection between emission intensity and spectral shape
while accounting for source-distance effects, we derived the phase-resolved
isotropic $\gamma$-ray luminosity. We find a statistically significant positive
correlation between $\log_{10}(L_{\gamma})$ and $\log_{10}(E_{\rm cut})$,
which can be described by $L_{\gamma}\propto E_{\rm cut}^{\alpha}$ with a
best-fit slope of $\alpha=0.74^{+0.16}_{-0.14}$. This slope is substantially
shallower than the value of $\alpha=2.31^{+0.22}_{-0.25}$ reported for
field MSPs by \citet{lsx+26}, which is consistent with the $\alpha=16/7\simeq2.29$
scaling expected for curvature radiation in the ECS framework \citep{khk+19}.
The physical origin of the shallow relation found for our GC-MSP sample remains
unclear because of the limited sample size and possible systematic effects.
In the following discussion, we examine several factors that may contribute
to the difference between GC MSPs and field MSPs, including cluster-associated
emission, differences in sample properties, and variations in magnetospheric
conditions or emission geometry. These possibilities are not mutually exclusive,
and the available data do not allow us to determine which, if any, is primarily
responsible for the observed difference.

Several factors may contribute to the observed difference between the 
$L_{\gamma}$--$E_{\rm cut}$ relations of GC and field MSPs.
From a physical perspective, two broad possibilities may contribute to the shallow relation. First, cluster-associated emission, potentially arising from IC processes or unresolved MSP contributions, may increase the inferred luminosity in faint phase intervals and compress the dynamic range of $L_{\gamma}$, thereby potentially flattening the observed relation. Second, variations in intrinsic magnetospheric emission properties, together with contributions from multiple emission zones or radiation processes, may modify the simple curvature-radiation scaling. These effects are not necessarily specific to GC MSPs, and the present data are insufficient to establish their relative importance or determine whether they account for the shallow relation in our sample.
Finally, several observational and methodological limitations, together with possible physical complications, preclude a definitive interpretation of the shallow $L_{\gamma}$--$E_{\rm cut}$ relation obtained in this work. Observationally and methodologically, our initial sample contains only four GC MSPs, and the phase-resolved analysis is restricted to the three sources with significant off-pulse detections. Consequently, the measured slope of $\alpha=0.74^{+0.16}_{-0.14}$ cannot yet be regarded as representative of the general GC-MSP population. In addition, the limited photon statistics in narrow phase bins and the angular resolution of \textit{Fermi}-LAT prevent a clean separation between intrinsic magnetospheric emission and possible cluster-associated components, including inverse-Compton (IC) emission from relativistic particles interacting with cluster radiation fields or unresolved contributions from other MSPs. The use of isotropic luminosities, which neglects potentially phase-dependent beaming corrections, introduces an additional systematic uncertainty. We therefore regard the present GC-MSP sample as an interesting pilot sample that motivates a larger survey.

Future $\gamma$-ray facilities with improved sensitivity and angular resolution, such as the proposed Very Large Area Gamma-ray Space Telescope (VLAST; \citealt{fcg+22}), together with longer exposure and a larger sample of GC MSPs with phase-connected timing solutions, will be essential for testing the robustness and population dependence of the measured $L_{\gamma}$--$E_{\rm cut}$ slope. Such data will also permit spatial and multi-component spectral analyses to better distinguish magnetospheric emission from cluster-related contributions, helping to clarify whether the shallow relation reflects intrinsic pulsar-emission physics, observational or methodological effects, or environmental effects specific to GCs.

\begin{acknowledgments}
We thank the anonymous referee for the helpful comments and suggestions.
This work was supported in part by the National Natural Science Foundation of China
under Grant Nos. 12233006 and 12103046, and by the Scientific Research and Innovation
Project for Postgraduate Students Pursuing Academic Degrees at Yunnan University under
Grant No. KC-262615807.
P.Z. and S.C. acknowledge support from the Xingdian Talent Support Plan--Youth Project.
\end{acknowledgments}

\bibliographystyle{aasjournalv7}
%\bibliography{ref}
\bibliography{aas}

\end{document}